# MOVPE growth of nitrogen- and aluminum-polar AlN on 4H-SiC

J. Lemettinen [a,*], H. Okumura [b,c], I. Kim [a], M. Rudzinski [d], J. Grzonka [d], T. Palacios [c], S. Suihkonen [a]

∗ Corresponding author
Email address: jori.lemettinen@aalto.fi (J. Lemettinen)

a) Department of Electronics and Nanoengineering, Aalto University, P.O BOX 13500, FIN-00076 AALTO, Finland

b) Faculty of pure and applied science, University of Tsukuba, Tsukuba 305-8573 Japan

c) Department of Electrical Engineering and Computer Science, Massachusetts Institute of Technology, Cambridge, MA 02139 USA

d) Institute of Electronic Materials Technology, Epitaxy Department (Z-15), Wolczynska Street 133, 01-919 Warsaw, Poland

**Abstract**
We present a comprehensive study on metal-organic vapor phase epitaxy growth of N-polar and Al-polar AlN on 4H-SiC with 4° miscut using constant growth parameters. At a high temperature of 1165°C, N-polar AlN layers had high crystalline quality whereas the Al-polar AlN surfaces had a high density of etch pits. For N-polar AlN, the V/III ratio below 1000 forms hexagonal hillocks, while the V/III ratio over 1000 yields step bunching without the hillocks. 1-µm-thick N-polar AlN layer grown in optimal conditions exhibited FWHMs of 307, 330 and 337 arcsec for (002), (102) and (201) reflections, respectively.

## 1. Introduction

III-nitride (III-N) alloys are currently used in a plethora of electronic and light emitting applications, such as, high-electron mobility transistors (HEMTs) and light-emitting diodes (LEDs). In III-Ns, AlN is a promising material for high-power and high-temperature applications because of its high critical electric field and high thermal conductivity. AlN is grown on sapphire, Si, silicon carbide (SiC), and native substrates [1–8].

SiC substrates are suitable for commercial high crystalline-quality AlN growth because of the small lattice mismatch of 1% between AlN and SiC in addition to a sufficiently large wafer size. AlN is typically grown by metal-organic vapor-phase epitaxy (MOVPE) and molecular beam epitaxy (MBE). MOVPE is preferred for commercial scale fabrication of AlN. The AlN layers grown on SiC by MOVPE have exhibited small X-ray rocking curve (XRC) full width half maximums (FWHMs) of 76 and 360 arcsec measured for (002) and (102) reflections, respectively [3].

Wurtzite III-N has two types of polar planes, metal-polar (001) and N-polar (00$\bar{1}$). N-polar III-Ns have several benefits for device applications due to the polarization induced electric field. For example, in N-polar GaN/AlGaN HEMTs the top GaN layers offer low contact resistances and the

AlGaN back barrier provides low buffer-leak current [9, 10]. N-polar III-Ns enable a more efficient carrier injection [9] in LEDs. In addition, N-polar III-Ns are used for laterally polarized structures with a large nonlinear susceptibility and polarization field along the lateral direction [11].

However, high-quality N-polar III-N growth is difficult because of higher binding energies, different diffusion pathways of adatoms and anisotropic lateral growth rate of surface steps, compared to metal-polar III-N growth [12, 13]. Especially, the low migration length of adatoms causes the formation of a high density of hillocks, leading to rough surfaces [14]. The migration length of adatoms depends on the growth temperature and V/III ratio. High growth temperatures decrease the surface roughness [7]. N-polar III-N layers are grown under V/III-ratios between 650 and 27500 [7, 14–17]. Additionally, the hillock formation can be suppressed by employing a substrate with a miscut of 4° together with a low growth rate [5, 7]. However, there are no reports of device quality N-polar AlN layers grown by MOVPE. Achievement of high-quality N-polar AlN growth by MOVPE would widen the range of optical and electrical device applications.

In this report, we present a comprehensive study on the MOVPE growth for both N-polar and Al-polar AlN on 4H-SiC with a miscut of 4°. The growth rate was kept at a constant low value by altering the precursor mass flows when either the V/III-ratio or temperature was varied to separate these effects on surface morphology and crystalline quality. The layer crystalline quality was evaluated using high-resolution x-ray diffraction (HR-XRD) measurements. The surface roughness was determined with atomic force microscopy (AFM). Defect selective etching was applied to confirm the sample polarity.

## 2. Experiment

200-nm-thick N- and Al-polar AlN layers were grown on C- and Si-face of 4H-SiC substrates, respectively, by MOVPE. The SiC substrates had an intentional miscut of 4° towards $<1\bar{1}0>$. The MOVPE reactor had a close-coupled showerhead configuration. Trimethylaluminium (TMAl) and ammonia ($NH_3$) were used as precursors for aluminum and nitrogen, respectively. Hydrogen was used as the carrier gas. The SiC surfaces were cleaned in-situ for 20 min at 1180°C in a hydrogen ambient prior to AlN growth. The nominal substrate surface temperature was between 1070°C and 1165°C during AlN growth. The temperature was estimated using an emissivity corrected in-situ pyrometer. The growth pressure was 50 mbar. The AlN layers were grown in a single step.

Three-axis HR-XRD measurements (Philips X'Pert Pro) were performed to assess the crystalline quality of the grown AlN layers. Rocking curve scans around the symmetrical (002) and various skew-symmetrical reflections were recorded. The setup consisted of an X-ray mirror, a 4 × Ge (220) monochromator and an analyzer crystal. X-ray wavelength of the copper $K_{\alpha1}$ emission line was used. Atomic force microscopy (AFM) (NT-MDT NTEGRA Aura) was used to measure the surface roughness of the grown AlN layers. Defect selective etching was applied to investigate the sample polarity [18–20]. The AlN layer surface was etched with either a molten eutectic mixture of KOH-NaOH with 10% of MgO powder or a 1M 50°C KOH solution and the layers were studied using scanning electron microscope (SEM).

## 3. Results and discussion

Common growth parameters for Al-polar AlN on Si-face SiC were used initially. N-polar and Al-polar AlN layers were grown at substrate surface temperature of 1070°C with a growth rate of 0.5 μm/h. Both N-polar and Al-polar AlN layers exhibited the same growth rate for the same growth conditions. Figure 1 shows AFM images of N-polar and Al-polar AlN layers with 200 nm thickness. The Al-polar layer had a quite smooth surface with few pits (RMS roughness: 1.46 nm). Conversely, the N-polar layer had a rough surface (RMS roughness: 10.1 nm). There were hexagonal hillocks with symmetrical structure, despite growth on 4° miscut substrates. The hexagonal hillock density was $1.8 \times 10^9$ /cm$^2$. The hillocks had no spiral shape, which is typically generated by screw/mixed dislocations and spiral growth [21]. We believe that multiple islands are formed on surfaces due to the small migration length of adatoms in N-polar plane. The enhancement of adatom mobility on surfaces is necessary to suppress the formation of hexagonal hillocks.

To study the N-polar and Al-polar AlN growth characteristics, a series of growths over a wide V/III-ratio range from 100 to 27000 were performed. A low growth rate of 0.1 μm/h and a high substrate-surface temperature of 1150°C, both of which enable the enhancement of the migration length of adatoms, were selected to suppress the formation of hexagonal hillocks. The growth parameters were adjusted towards values reported for N-polar AlN growth because Al-polar AlN has a wider growth window for good crystalline quality [3, 7, 22]. The AlN layers were grown in a single step. In case of AlN on SiC growth, changing the growth conditions after nucleation has a minor effect on film quality[23, 24]. AlN layers with different polarity were grown simultaneously at the same V/III-ratio. The growth rate decreases with increasing the V/III-ratio due to parasitic reactions. In order to separate the effects of V/III-ratio and growth rate, the growth rate of the AlN layers was maintained at 0.1 μm/h throughout the series by altering the precursor mass flow.

Figure 2 shows AFM images of N-polar (a, c and e) and Al-polar (b, d and f) AlN layers with 200 nm thickness. The N- and Al-polar AlN layer morphology is considerably different. The N-polar AlN surfaces show pronounced step bunching towards the substrate miscut marked by arrows in Figure 2. The step bunching may be mitigated by nucleating a lower temperature layer, such as N-polar GaN,
grown on the SiC substrate [25]. There were no multi-micron sized hexagonal hillocks which are attributed to the existence of inversion domains [7, 13]. In contrast, the Al-polar AlN surfaces had a step-and-terrace structure without large step bunching but there are pits, whose density decrease with increasing V/III-ratio.

The V/III-ratio has a significant impact on the dominating growth mode of N-polar AlN as seen from Figures 2 a, c and e. As shown in Figure 2a, sub-micron sized hexagonal hillocks formed on the N-polar AlN surface grown at a very low V/III-ratio of 100. At the high V/III-ratio of 1000 (Figure 2c), the step-flow growth mode becomes dominant, suppressing the formation of hexagonal hillocks and yielding step bunching. The steps advance toward the substrate miscut direction of < 1̄10 >. The average step height was 7 nm, or 28 monolayers of AlN. Further increase of the V/III-ratio to 27000 (Figure 2e) does not significantly affect the surface roughness. We consider that the step bunching is caused by anisotropy in the terrace step adatom capture probabilities [26]. The anisotropy in capture probability leads to differences in step lateral advancement rates and can cause step-bunching. The step bunching is more prominent under growth conditions where the adatom migration length is greater than the surface terrace width [27].

The dependence of surface kinetics on vapor supersaturation has been used to explain the surface morphology of Al-polar AlN [27]. The surface-kinetics model predicts that increasing the growth temperature and lowering the growth rate can reduce the surface supersaturation, and thus, reduce the island formation, consistent with our results. Additionally, increasing the V/III-ratio in this growth window shows the elimination of hillocks and the onset of step-bunching. However, it is difficult to estimate the dependence surface supersaturation on V/III-ratio due to the influence of other process parameters. Further investigation on the relation between V/III-ratio and supersaturation is necessary. Increasing the V/III-ratio has decreased hillock formation in N-polar AlN [7]. Conversely, in Al-polar AlN growth, decreasing the V/III-ratio can initiate step-bunching [26]. This difference could be explained by the differences in surface reconstruction and diffusion barriers between N- and Al-polar AlN [26].

For Al-polar AlN growth, a high density of pits (5.0 × $10^{10}$ /cm$^2$) was observed on the sample surface grown at the low V/III-ratio of 100 (Figure 2b). These pits are generated around dislocations due to the etching effect of the H$_2$ carrier gas [28]. The pit density significantly decreases to 8.8 × $10^9$ /cm$^2$ as the V/III-ratio increases to 1000 (Figure 2d). High V/III-ratio of 27000 (Figure 2f) reduces the pit density to 7.5 × $10^7$ /cm$^2$. We suggest that the low growth rate at high growth temperature causes nitrogen desorption from the growth surface and around edge dislocations, forming pits. A high V/III-ratio, or high nitrogen partial pressure, may be effective in suppressing the nitrogen desorption.

Defect selective etching was applied to confirm the sample polarity. The Al-polar and N-polar AlN layers grown at a nominal substrate surface temperature of 1150°C with a V/III-ratio of 1000 and 27000 were etched with a molten eutectic mixture of KOH-NaOH with 10% of MgO (E+M) powder. Figure 3 presents the etching results from samples grown using a V/III-ratio of 1000. Identical results were obtained for the other sample pair (not shown).

Wet etching is an accurate method for determining III-N layer polarity due to the vastly different etch rates of N-polar and Al-polar surfaces [20, 29]. Due to very thin 200 nm layers the Al-polar was only etched for 30 seconds and for N-polar AlN layers the etching time was reduced down to 2 seconds and still the layer was almost completely dissolved (Figure 3b). Therefore to lower the etch rate, N-polar samples were additionally etched in 1M 50°C KOH solution which resulted in a typical N-polar hexagonal pyramids (Figure 3c) which are attributed to being bound by more chemically stable crystalline facets [29, 30]. Al-polar AlN layer exhibited typical defect selective etch behavior (Figure 3a) where the hexagonal pits are generated on dislocation cores [18, 19]. In our case hexagonal pits were already generated during growth (Figure 2) due to the in-situ etching effect of the H$_2$ carrier and after the second etching in KOH-NaOH they were enlarged. The pit density (8.8×$10^9$ /cm$^2$) counted before etching is in good agreement with the hexagonal pit density (1.2 × $10^{10}$ /cm$^2$) after KOH-NaOH etching implying that the surface pits result from threading dislocations.

Figure 4 presents XRC FWHM for the grown N-polar and Al-polar AlN layers. The FWHMs were calculated using least-squares fitting of a Gaussian line-shape. Rocking curve scans around the symmetrical (002) and skew-symmetrical (102) reflections were recorded. Only screw dislocation densities contribute to the broadening of the (002) peak, while both screw and edge dislocations contribute to the broadening of the (102) peak [31]. Typically the density of edge dislocations is larger than that of screw dislocations in III-N layers. Figure 5 presents an XRC (002) line-shape recorded from a 200-nm-thick Al-polar AlN layer grown at 1150°C with a V/III-ratio of 27000. Thin epitaxial III-N layers can exhibit narrow specular diffraction in addition to broad diffuse scattering [32–34]. The narrow specular peak is attributed to a high density of inversion domains [32] or correlated dislocations [33]. Gaussian and Lorentzian line-shapes have been fitted to the experimental data with FWHMs of 314 arcsec and 7 arcsec, respectively. The FWHM of the Lorentzian peak is at the limit of instrument resolution. In Figure 4 the FWHM of the diffuse scattering was used.

Based on Figure 4, for N-polar AlN, the minimum values of (002) FWHM was achieved at a V/III-ratio of 1000, while the (102) FWHM does not show a clear trend with the V/III-ratio. The V/III-ratio of 1000 shows the smallest FWHMs of 232 and 509 arcsec for the (002) and (102) scan, respectively. We consider that the high lateral ad-atom mobility at comparatively high V/III-ratio enhances the step-flow growth mode, improving crystalline quality. In contrast, the Al-polar AlN layers have a vastly different optimum V/III-ratio compared to the N-polar AlN layers. As shown in Figure 4, for Al-polar AlN, the highest V/III-ratio of 27000 resulted in the smallest FWHMs, indicating the highest crystalline quality for this sample series. The FWHMs for (002) and (102) reflections were 314 and 576 arcsec, respectively. The FWHM of the (002) scan has a clear decreasing trend with increasing V/III-ratio. The (102) scan FWHM does not have as clear a trend as the (002) FWHM. For Al-polar samples grown with V/III-ratio below 1000, the (102) peak width was too broad to accurately calculate the FWHM.

To investigate the effect of growth temperature on N-polar and Al-polar AlN growth, both N-polar and Al-polar AlN layers were grown simultaneously using a V/III-ratio of 1000 and nominal substrate surface temperatures between 1080°C and 1165°C. The growth rate of both N-polar and Al-polar AlN layers was maintained at 0.1 μm/h throughout the series by altering the precursor mass flows. N-polar AlN surfaces are independent of the growth temperature, while Al-polar AlN surfaces have a high density of pits as the growth temperature increases, e.g., $5.0 \times 10^{10}$ /cm$^2$ at 1150°C. Figure 6 shows the XRC FWHMs for N-polar and Al-polar AlN layers for this series. For N-polar AlN layers, the XRC FWHMs decrease with increasing growth temperature, indicating improvement of crystalline quality. At the high growth temperature of 1165°C, we achieved the lowest XRC FWHMs of 220 and 514 arcsec for (002) and (102) reflections, respectively. In contrast, for Al-polar AlN layers, there is no clear trend between the growth temperature and crystalline quality, indicating that the crystalline quality of Al-polar AlN layers is limited by the V/III-ratio in this growth regime. For Al-polar AlN grown with two step approach with V/III-ratios of 2338 and 116, XRC FWHMs of 76 and 360 arcsec were reported for (002) and (102) reflections, respectively [3]. Further improvement of the crystalline quality for Al-polar AlN growth is expected by optimizing the V/III ratio and growth rate.

A 1-μm-thick N-polar AlN layer was grown in order to confirm the stability of the step flow growth mode and to allow more accurate crystalline quality analysis. Figure 7 shows an AFM image of the N-polar AlN layer grown using a V/III-ratio of 1000 and a nominal substrate surface temperature of 1150°C. The growth rate was 0.1 μm/h. Compared to a 200-nm-thick N-polar AlN layer with same growth conditions (Figure 2 c), the thicker AlN layer surface has a lower density of steps with a greater of height around 40 nm. Because the step-bunching effect is not self-limiting the surface roughness increases with increasing epitaxial layer thickness [5].

Figure 8 presents a series of XRC FWHMs as a function of the modified inclination angle $\sin^2\chi$. All the scans were recorded in a skew-symmetrical three-axis configuration. The FWHMs of the skew-symmetrical scans, namely the 105 scan, have a smaller minimum value than the symmetrical 002 scan. This could be due to anisotropic stress created by the substrate miscut or the difference in symmetry between the scans [35]. The contributions of crystalline tilt and twist to the XRC peak broadening, assuming high dislocation density and Gaussian peak shape, are connected through following expression [36]:

$$\beta_{hkl}^2 = \left(\beta_{tilt}\cos\chi\right)^2 + \left(\beta_{twist}\sin\chi\right)^2, \tag{1}$$

where β hkl , β tilt , β twist and χ are the measured width of a diffraction peak with Miller indices (hkl), peak broadening due to crystalline tilt, peak broadening due to crystalline twist and measurement inclination angle for the diffraction peak with Miller indices (hkl), respectively. Equation 1 can be converted to be a function of $\sin^2\chi$ for linear fitting:

$$\beta_{hkl}^2 = \left(\beta_{tilt}^2 - \beta_{twist}^2\right)\sin\chi^2 + \beta_{tilt}^2. \tag{2}$$

Typically the reflection having the higher inclination angle is considered to be a better figure of merit for III-N overall crystalline quality due to the higher density of edge dislocations. The reported XRC FWHMs of N-polar AlN layers grown on SiC by MOVPE are 468 and 684 arcsec for

(002) and (102) reflections, respectively [14]. The FWHMs of of N-polar GaN layers grown on SiC by MOVPE are 108 and 396 arcsec for (002) and (201) reflections [37]. In comparison, our 1-μm-thick N-polar AlN layer exhibited quite small FWHMs of 307, 330 and 337 arcsec for (002), (102) and (201) reflections due to a thick layer and high crystalline quality. The (002) FWHM of the 1-μm-thick N-polar AlN layer was higher than that of the 200-nm-thick N-polar AlN layer grown with same parameters. The increase in peak width could be due to the increased surface roughness induced by step bunching.

The obtained peak broadening components can be used to estimate the screw and edge dislocation densities, assuming randomly distributed dislocations, using the following expressions [38]:

$$\rho_{screw} = \frac{\Gamma_{tilt}^2}{4.36\, b_{screw}^2}, \quad (3)$$

$$\rho_{edge} = \frac{\Gamma_{twist}^2}{4.36\, b_{edge}^2}, \quad (4)$$

where ρ, Γ and b are the dislocation density, the FWHM corresponding to the dislocation type expressed in radians and the length of the Burgers vector, respectively. The Burgers vectors for AlN are $b_{screw}$ = 4.98 Å (along [001]) and $b_{edge}$ = 3.11 Å (1/3 along [110]) [31]. The obtained dislocation density estimates for the 1-μm-thick AlN layer are $\rho_{screw}$ = 2 × $10^8$ /cm$^2$ and $\rho_{edge}$ = 7 × $10^8$ /cm$^2$. As expected, the edge dislocation density dominates the crystalline quality, which is typical behavior for III-Ns. The threading dislocation density (TDD) is 9 × $10^8$ /cm$^2$. The reported lowest TDD for N-polar GaN is 2.1 × $10^9$ /cm$^2$ [39, 40]. For a 200-nm-thick N-polar AlN grown by MBE on SiC the estimated TDD was 2.1 × $10^9$ /cm$^2$ [41]. We conclude that the crystalline quality of our N-polar AlN layers is slightly better than that reported for N-polar III-Ns.

**4. Conclusions**

N- and Al-polar AlN layers were grown on C- and Si-surfaces of 4H-SiC substrates with an intentional miscut of 4° towards < 1̄10 > by MOVPE. A low growth rate of 0.1 μm/h was effective for N-polar AlN to improve the crystalline quality and to suppress hexagonal hillocks. At a high temperature of 1165°C, N-polar AlN layers had high crystalline quality but showed a step bunching surface morphology, whereas the Al-polar AlN surfaces had a high density of etch pits. For N-polar AlN, the V/III ratio below 1000 forms hexagonal hillocks, while the V/III ratio over 1000 yields step bunching without the hillocks. 1-μm-thick N-polar AlN layer grown in optimal conditions exhibited FWHMs of 307, 330 and 337 arcsec for (002), (102) and (201) reflections, respectively. Estimated dislocation densities for the N-polar AlN layer are $\rho_{screw}$ = 2 × $10^8$ /cm$^2$ and $\rho_{edge}$ = 7 × $10^8$ /cm$^2$. This crystalline quality is slightly better than that of reported N-polar III-Ns.

**Acknowledgements**

This work was supported by the Academy of Finland (grant 297916), the Foundation for Aalto University Science and Technology, JSPS KAKENHI Grant No. 15H06070 and 16H06424 and National Centre for Research and Development in the frame of projects PBS3/A3/23/2015. A part of the research was performed at the OtaNano - Micronova Nanofabrication Centre of Aalto University.

# References

[1] M. Sakai, H. Ishikawa, T. Egawa, T. Jimbo, M. Umeno, T. Shibata, K. Asai, S. Sumiya, Y. Kuraoka, M. Tanaka, O. Oda, Growth of high-quality GaN films on epitaxial AlN/sapphire templates by MOVPE, Journal of Crystal Growth 244 (1) (2002) 6–11. doi:10.1016/S0022-0248(02)01573-7.

[2] H. Lahrèche, P. Vennéguès, O. Tottereau, M. Laügt, P. Lorenzini, M. Leroux, B. Beaumont, P. Gibart, Optimisation of AlN and GaN growth by metalorganic vapour-phase epitaxy (MOVPE) on Si(111), Journal of Crystal Growth 217 (1) (2000) 13–25. doi:10.1016/S0022-0248(00)00478-4.

[3] M. Imura, H. Sugimura, N. Okada, M. Iwaya, S. Kamiyama, H. Amano, I. Akasaki, A. Bandoh, Impact of high-temperature growth by metalorganic vapor phase epitaxy on microstructure of AlN on 6H-SiC substrates, Journal of Crystal Growth 310 (7-9) (2008) 2308–2313. doi:10.1016/j.jcrysgro.2007.11.206.

[4] H. J. Kim, S. Choi, D. Yoo, J. H. Ryou, R. D. Dupuis, R. F. Dalmau, P. Lu, Z. Sitar, Modulated precursor flow epitaxial growth of AlN layers on native AlN substrates by metal-organic chemical vapor deposition, Applied Physics Letters 93 (2) (2008) 2006–2009. doi:10.1063/1.2959064.

[5] S. Keller, N. A. Fichtenbaum, F. Wu, D. Brown, A. Rosales, S. P. Denbaars, J. S. Speck, U. K. Mishra, Influence of the substrate misorientation on the properties of N-polar GaN films grown by metal organic chemical vapor deposition, Journal of Applied Physics 102 (8). doi:10.1063/1.2801406.

[6] S. Keller, Y. Dora, F. Wu, X. Chen, S. Chowdury, S. P. Denbaars, J. S. Speck, U. K. Mishra, Properties of N-polar GaN films and AlGaN/GaN heterostructures grown on (111) silicon by metal organic chemical vapor deposition, Applied Physics Letters 97 (14) (2010) 1–4. doi:10.1063/1.3499428.

[7] D. Won, J. M. Redwing, Effect of AlN buffer layers on the surface morphology and structural properties of N-polar GaN films grown on vicinal C-face SiC substrates, Journal of Crystal Growth 377 (2013) 51–58. doi:10.1016/j.jcrysgro.2013.04.038. URL http://dx.doi.org/10.1016/j.jcrysgro.2013.04.038

[8] J. Weyher, P. Brown, A. Zauner, S. Müller, C. Boothroyd, D. Foord, P. Hageman, C. Humphreys, P. Larsen, I. Grzegory, S. Porowski, Morphological and structural characteristics of homoepitaxial GaN grown by metalorganic chemical vapour deposition (MOCVD), Journal of Crystal Growth 204 (4) (1999) 419–428. doi:10.1016/S0022-0248(99)00217-1.

[9] S. Keller, H. Li, M. Laurent, Y. Hu, N. Pfaff, J. Lu, D. F. Brown, N. a. Fichtenbaum, J. S. Speck, S. P. Denbaars, U. K. Mishra, Recent progress in metal-organic chemical vapor deposition of (000-1) N-polar group-III nitrides, Semiconductor Science and Technology 29 (2014) 113001. doi:10.1088/0268-1242/29/11/113001.

[10] M. H. Wong, S. Keller, N. Dasgupta, D. J. Denninghoff, S. Kolluri, D. F. Brown, J. Lu, N. a. Fichtenbaum, E. Ahmadi, U. Singisetti, A. Chini, S. Rajan, S. P. DenBaars, J. S. Speck, U. K. Mishra, N-polar GaN epitaxy and high electron mobility transistors, Semiconductor Science and Technology 28 (7) (2013) 74009. doi:10.1088/0268-1242/28/7/074009. URL http://iopscience.iop.org/0268-1242/28/7/074009/article/

[11] R. Katayama, Y. Kuge, T. Kondo, K. Onabe, Fabrication of lateral lattice-polarity-inverted GaN heterostructure, Journal of Crystal Growth 301 (2007) 447–451.

[12] T. K. Zywietz, J. Neugebauer, M. Scheffler, The adsorption of oxygen at GaN surfaces, Applied Physics Letters 74 (12) (1999) 1695. doi:10.1063/1.123658. URL http://scitation.aip.org/content/aip/journal/apl/74/12/10.1063/1.123658

[13] A. R. A. Zauner, E. Aret, W. J. P. Van Enckevort, J. L. Weyher, S. Porowski, J. J. Schermer, Homo-epitaxial growth on the N-face of GaN single crystals: The influence of the misorientation on the surface morphology, Journal of Crystal Growth 240 (1-2) (2002) 14–21. doi:10.1016/S0022-0248(01)02389-2.

[14] S. Keller, N. Fichtenbaum, F. Wu, G. Lee, S. P. Denbaars, J. S. Speck, U. K. Mishra, Effect of the nucleation conditions on the polarity of AlN and GaN films grown on C-face 6H-SiC, Japanese Journal of Applied Physics, Part 2: Letters 45 (8-11). doi:10.1143/JJAP.45.L322.

[15] T. Matsuoka, Y. Kobayashi, H. Takahata, T. Mitate, S. Mizuno, A. Sasaki, M. Yoshimoto, T. Ohnishi, M. Sumiya, N-polarity GaN on sapphire substrate grown by MOVPE, Physica Status Solidi (B) Basic Research 243 (7) (2006) 1446–1450. doi:10.1002/pssb.200565456.

[16] S. Keller, C. S. Suh, Z. Chen, R. Chu, S. Rajan, N. A. Fichtenbaum, M. Furukawa, S. P. DenBaars, J. S. Speck, U. K. Mishra, Properties of N-polar AlGaN/GaN heterostructures and field effect transistors grown by metalorganic chemical vapor deposition, Journal of Applied Physics 103 (3) (2008) 1–5. doi:10.1063/1.2838214.

[17] D. Won, X. Weng, Z. Y. Al Balushi, J. M. Redwing, Influence of growth stress on the surface morphology of N-polar GaN films grown on vicinal C-face SiC substrates, Applied Physics Letters 103 (24). doi:10.1063/1.4845575.

[18] J. L. Weyher, S. Lazar, L. Macht, Z. Liliental-Weber, R. J. Molnar, S. Müller, V. G. M. Sivel, G. Nowak, I. Grzegory, Orthodox etching of HVPE-grown GaN, Journal of Crystal Growth 305 (2 SPEC. ISS.) (2007) 384–392. doi:10.1016/j.jcrysgro.2007.03.030.

[19] J. L. Weyher, Characterization of wide-band-gap semiconductors (GaN,SiC) by defect-selective etching and complementary methods, Superlattices and Microstructures 40 (4-6 SPEC. ISS.) (2006) 279–288. doi:10.1016/j.spmi.2006.06.011.

[20] J. L. Rouviere, J. L. Weyher, M. Seelmann-Eggebert, S. Porowski, Polarity determination for GaN films grown on (0001) sapphire and high-pressure-grown GaN single crystals, Applied Physics Letters 73 (5) (1998) 668–670. doi:10.1063/1.121942.

[21] T. Akasaka, Y. Kobayashi, M. Kasu, Nucleus and spiral growth mechanisms of GaN studied by using selective-area metalorganic vapor phase epitaxy, Applied physics express 3 (7) (2010) 075602.

[22] D. F. Brown, R. Chu, S. Keller, S. P. Denbaars, U. K. Mishra, Electrical properties of N-polar AlGaN/GaN high electron mobility transistors grown on SiC by metalorganic chemical vapor deposition, Applied Physics Letters 94 (15) (2009) 148–151. doi:10.1063/1.3122347.

[23] S. Kitagawa, H. Miyake, K. Hiramatsu, High-quality aln growth on 6H-SiC substrate using three dimensional nucleation by low-pressure hydride vapor phase epitaxy, Japanese Journal of Applied Physics 53 (5S1) (2014) 05FL03.

[24] H. Hirayama, S. Fujikawa, J. Norimatsu, T. Takano, K. Tsubaki, N. Kamata, Fabrication of a low threading dislocation density ELO-AlN template for application to deep-uv leds, physica status solidi (c) 6 (S2).

[25] D. F. Brown, S. Keller, F. Wu, J. S. Speck, S. P. DenBaars, U. K. Mishra, Growth and characterization of N-polar GaN films on SiC by metal organic chemical vapor deposition, Journal of Applied Physics 104 (2) (2008) 1–6. doi:10.1063/1.2956329.

[26] K. Bellmann, U. W. Pohl, C. Kuhn, T. Wernicke, M. Kneissl, Controlling the morphology transition between step-flow growth and step-bunching growth, Journal of Crystal Growth 478 (2017) 187–192.

[27] I. Bryan, Z. Bryan, S. Mita, A. Rice, J. Tweedie, R. Collazo, Z. Sitar, Surface kinetics in AlN growth: A universal model for the control of surface morphology in III-nitrides, Journal of Crystal Growth 438 (2016) 81–89.

[28] J. Bai, T. Wang, P. J. Parbrook, I. M. Ross, A. G. Cullis, V-shaped pits formed at the GaN/AlN interface, Journal of Crystal Growth 289 (1) (2006) 63–67. doi:10.1016/j.jcrysgro.2005.10.146.

[29] Y. Fu, Q. Fan, S. Chevtchenko, Ü. Özgür, H. Morkoç, Y. Ke, R. Devaty, W. J. Choyke, C. K. Inoki, T. S. Kuan, Growth and polarity control of GaN and AlN on carbon-face SiC by metalorganic vapor phase epitaxy, Proceedings of SPIE 6473 (2007) 647305. doi:10.1117/12.706938. URL http://proceedings.spiedigitallibrary.org/proceeding.aspx?doi=10.1117/12.706938

[30] X. Ni, Ü. Özgür, H. Morkoç, Z. Liliental-Weber, H. O. Everitt, Epitaxial lateral overgrowth of a-plane GaN by metalorganic chemical vapor deposition, Journal of Applied Physics 102 (5) (2007) 053506. doi:10.1063/1.2773692. URL http://aip.scitation.org/doi/10.1063/1.2773692

[31] M. Moram, M. Vickers, X-ray diffraction of III-nitrides, Reports on Progress in Physics 72 (3) (2009) 036502.

[32] H. Heinke, V. Kirchner, H. Selke, R. Chierchia, R. Ebel, S. Einfeldt, D. Hommel, X-ray scattering from GaN epitaxial layers - an example of highly anisotropic coherence, Journal of Physics D: Applied Physics 34 (10A) (2001) A25–A29. doi:10.1088/0022-3727/34/10A/306.

[33] V. M. Kaganer, O. Brandt, H. Riechert, K. K. Sabelfeld, X-ray diffraction of epitaxial films with arbitrarily correlated dislocations: Monte Carlo calculation and experiment, Physical Review B - Condensed Matter and Materials Physics 80 (3) (2009) 1–4. doi:10.1103/PhysRevB.80.033306.

[34] Q. Zhu, A. Botchkarev, W. Kim, Ö. Aktas, A. Salvador, B. Sverdlov, H. Morkoç, S. C. Y. Tsen, D. J. Smith, Structural properties of GaN films grown on sapphire by molecular beam epitaxy, Applied Physics Letters 68 (8) (1996) 1141–1143. doi:10.1063/1.115739.

[35] H. Kim-Chauveau, P. De Mierry, H. Cabane, D. Gindhart, In-plane anisotropy characteristics of GaN epilayers grown on a-face sapphire substrates, Journal of Applied Physics 104 (11) (2008) 113516.

[36] S. R. Lee, A. M. West, A. A. Allerman, K. E. Waldrip, D. M. Follstaedt, P. P. Provencio, D. D. Koleske, C. R. Abernathy, Effect of threading dislocations on the bragg peakwidths of GaN, AlGaN, and AlN heterolayers, Applied Physics Letters 86 (24) (2005) 241904.

arXiv:http://dx.doi.org/10.1063/1.1947367, doi:10.1063/1.1947367. URL http://dx.doi.org/10.1063/1.1947367

[37] S. Kolluri, S. Member, S. Keller, S. P. Denbaars, U. K. Mishra, Microwave Power Performance N-Polar GaN MISHEMTs Grown by MOCVD on SiC Substrates Using an Al2O3 Etch-Stop Technology, IEEE Electron Device Letters 33 (1) (2012) 44–46.

[38] C. Dunn, E. Kogh, Comparison of dislocation densities of primary and secondary recrystallization grains of Si-Fe, Acta Metallurgica 5 (10) (1957) 548 – 554. doi:http://dx.doi.org/10.1016/0001-6160(57)90122-0. URL http://www.sciencedirect.com/science/article/pii/0001616057901220

[39] K. Kusakabe, K. Kishino, A. Kikuchi, T. Yamada, D. Sugihara, S. Nakamura, Reduction of threading dislocations in migration enhanced epitaxy grown GaN with N-polarity by use of AlN multiple interlayer, Journal of Crystal Growth 230 (3-4) (2001) 387–391. doi:10.1016/S0022-0248(01)01248-9.

[40] D. Won, X. Weng, J. M. Redwing, Effect of indium surfactant on stress relaxation by V-defect formation in GaN epilayers grown by metalorganic chemical vapor deposition, Journal of Applied Physics 108 (9). doi:10.1063/1.3487955.

[41] H. Okumura, T. Kimoto, J. Suda, Growth of nitrogen-polar 2H-AlN on step-height-controlled 6H-SiC(0001) substrate by molecular-beam epitaxy, Japanese Journal of Applied Physics 51 (2 PART 2). doi:10.1143/JJAP.51.02BH02.

**Figure captions**

Figure 1: 2 × 2 μm 2 AFM images of 200-nm-thick N- and Al-polar AlN layers grown on 4˚ off 4H-SiC substrates at 1070˚C nominal substrate surface temperature with V/III-ratio of 1000.

Figure 2: 2 × 2 μm 2 AFM images of 200-nm-thick N- and Al-polar AlN layers grown on 4˚ off 4H-SiC substrates at 1150˚C nominal substrate surface temperature with varying V/III-ratio. The substrate miscut direction is indicated by a white arrow.

Figure 3: SEM images of 200-nm-thick Al- and N-polar AlN layers after etching with: a molten eutectic mixture of KOH-NaOH with 10% of MgO powder (a and b), and a 1M 50˚C KOH solution (c).

Figure 4: XRC FWHMs of symmetrical (002) (filled symbols) and skew-symmetrical (102) (open symbols) scans for 200-nm-thick N-polar (black) and Al-polar (red) AlN layers grown on 4˚ off 4H-SiC substrates at 1150˚C nominal substrate surface temperature with varying V/III-ratio.

Figure 5: XRC scan from a 200-nm-thick Al-polar AlN layer (solid black) grown on 4˚ off 4H-SiC substrate at 1150˚C nominal substrate surface temperature with V/III-ratio of 27000. Gaussian and Lorentzian line-shapes fit to the XRC scan solid red and dotted black, respectively.

Figure 6: XRC FWHMs of symmetrical (002) (filled symbols) and skew-symmetrical (102) (open symbols) scans for 200-nm-thick for N-polar (black) and Al-polar (red) AlN layers grown on 4˚ off 4H-SiC substrates with V/III-ratio of 1000 at varying nominal substrate surface temperature.

Figure 7: 2 × 2 μm 2 AFM image of 1-μm-thick N-polar AlN layer grown on 4° off 4H-SiC substrate at 1150°C nominal substrate surface temperature with V/III-ratio of 1000. The substrate miscut direction is indicated by a white arrow.

Figure 8: XRC FWHMs of various skew-symmetrical scans for a 1-μm-thick N-polar AlN layer grown on 4° off 4H-SiC substrate at 1150°C nominal substrate surface temperature with V/III-ratio of 1000.

**Figures**

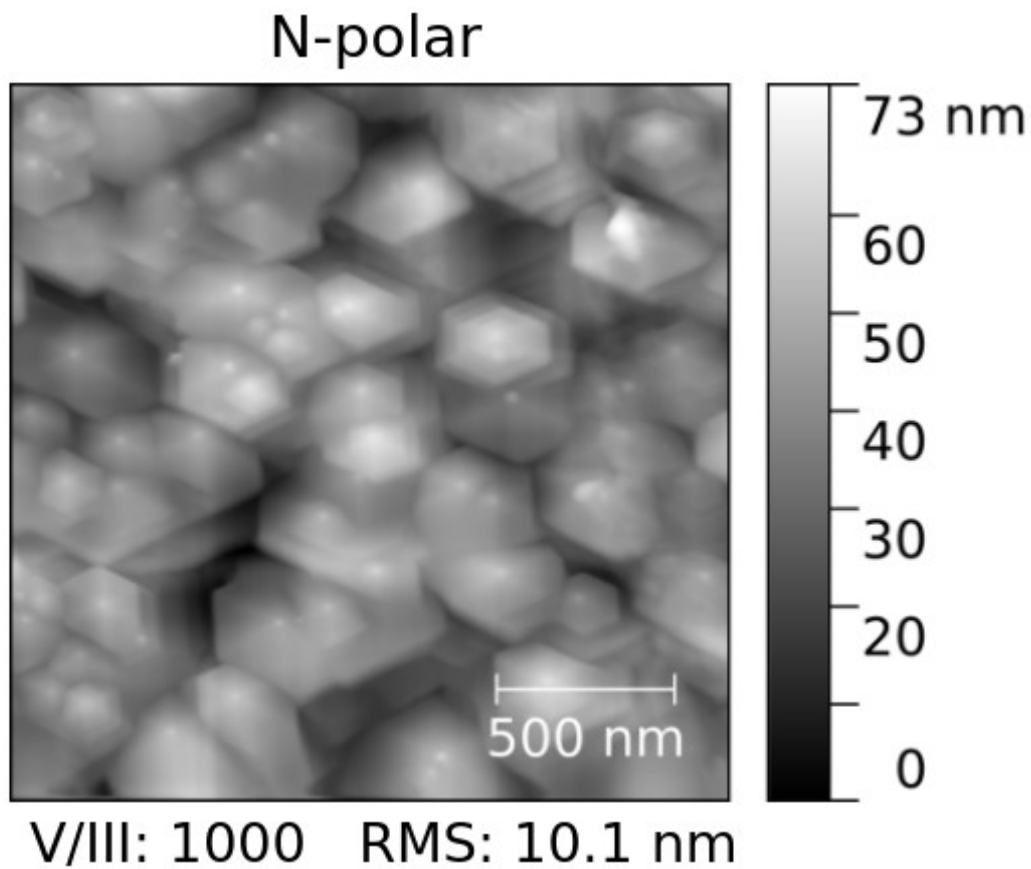

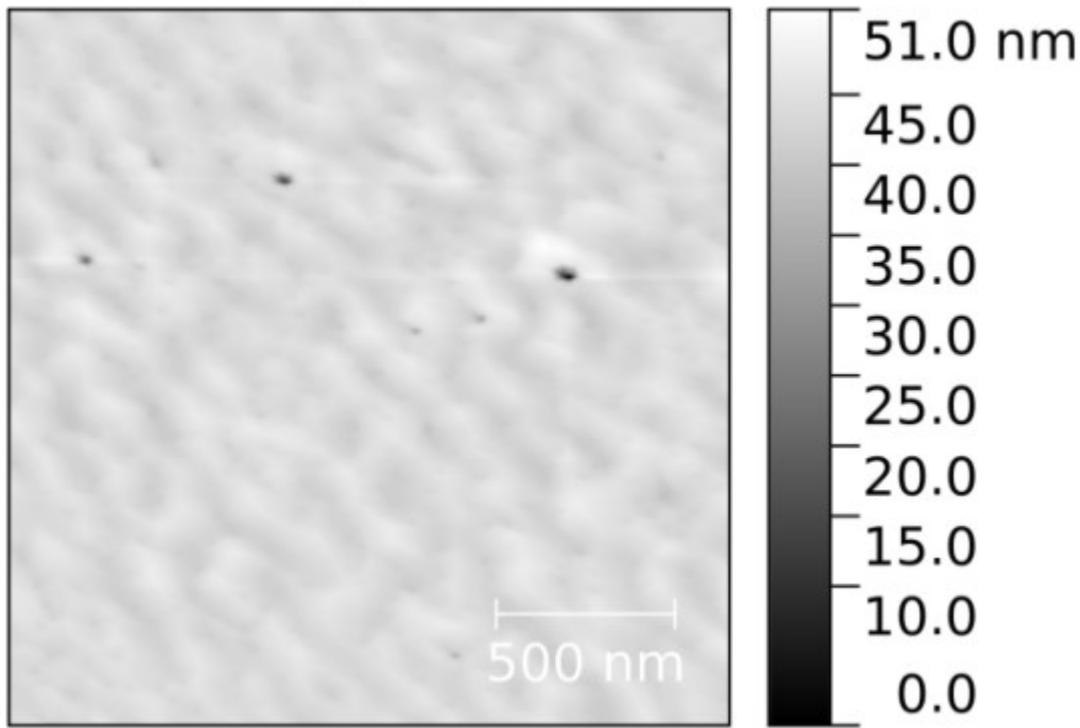

V/III: 1000    RMS: 1.46 nm

**Figure 1.**

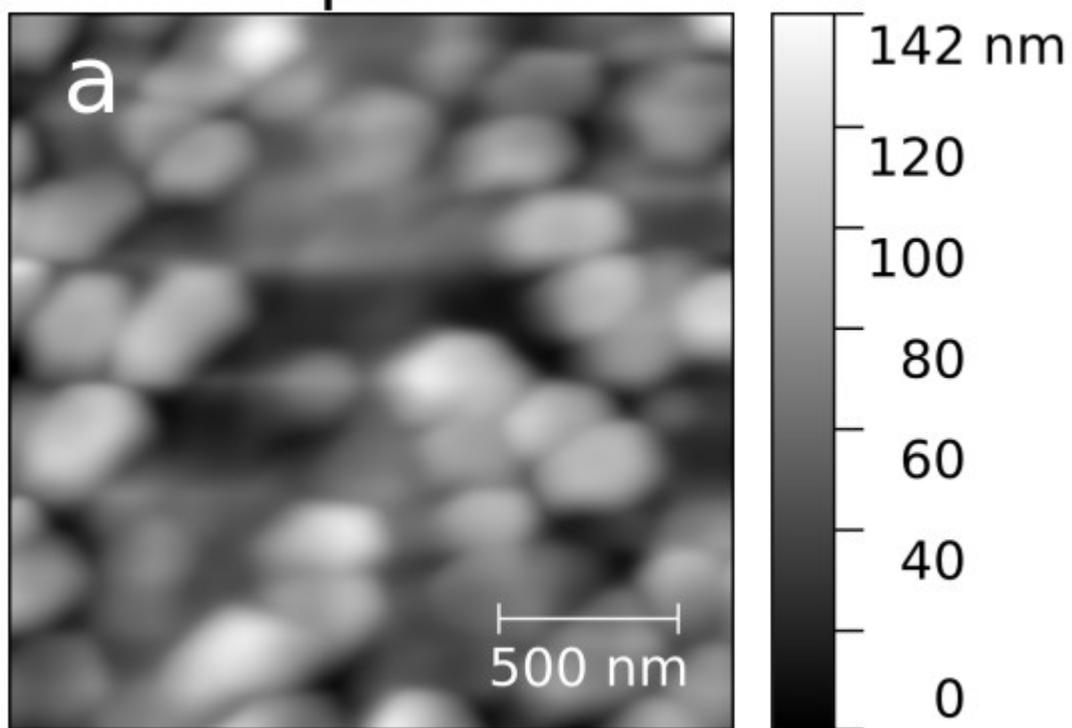

V/III: 100    RMS: 24.6 nm

## Al-polar

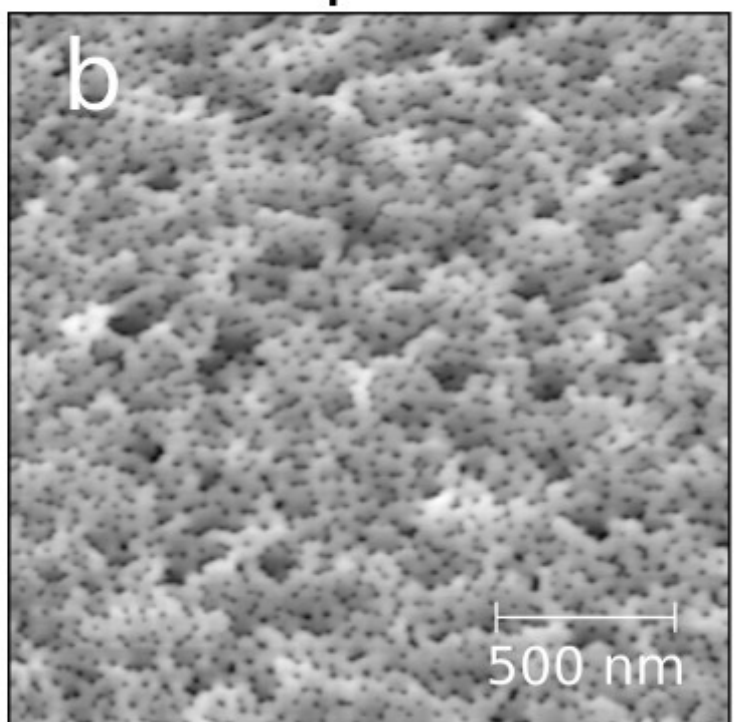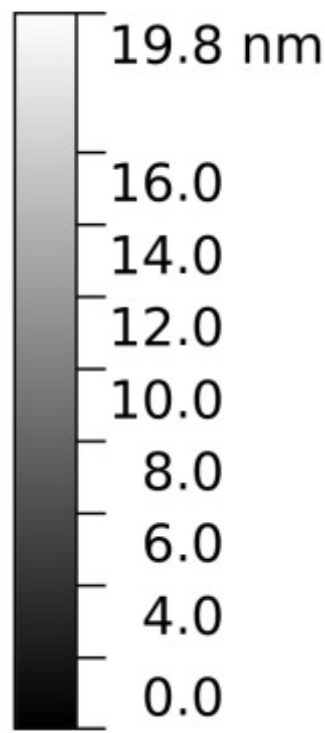

V/III: 100     RMS: 2.43 nm

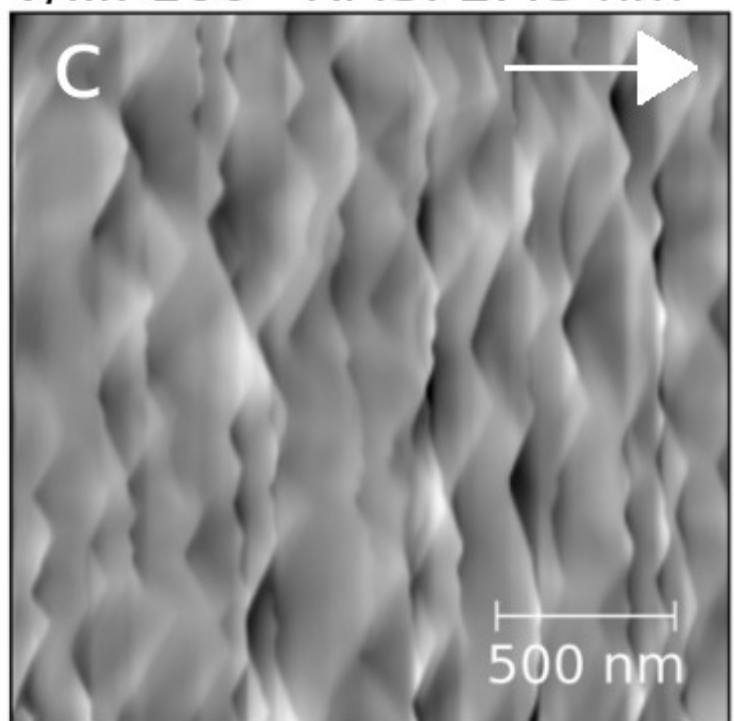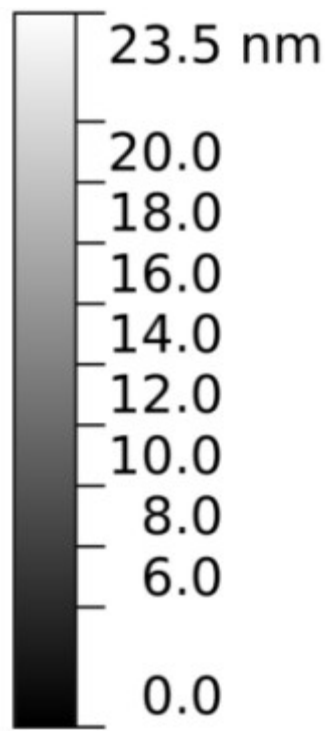

V/III: 1000    RMS: 2.92 nm

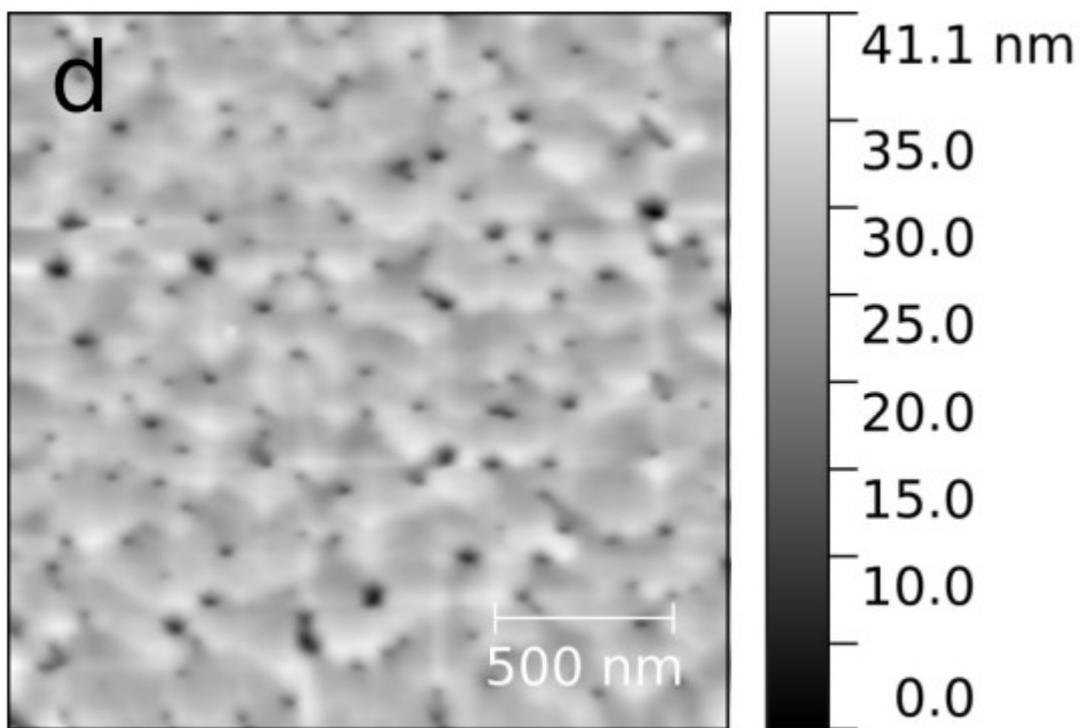

V/III: 1000    RMS: 3.62 nm

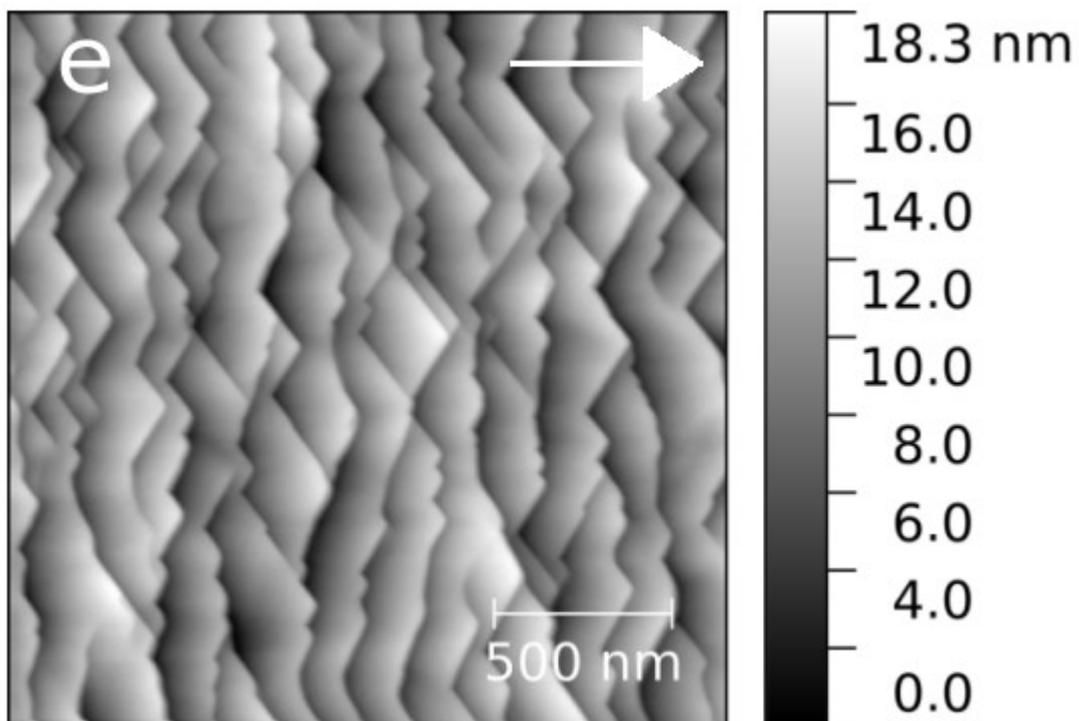

V/III: 27000    RMS: 2.69 nm

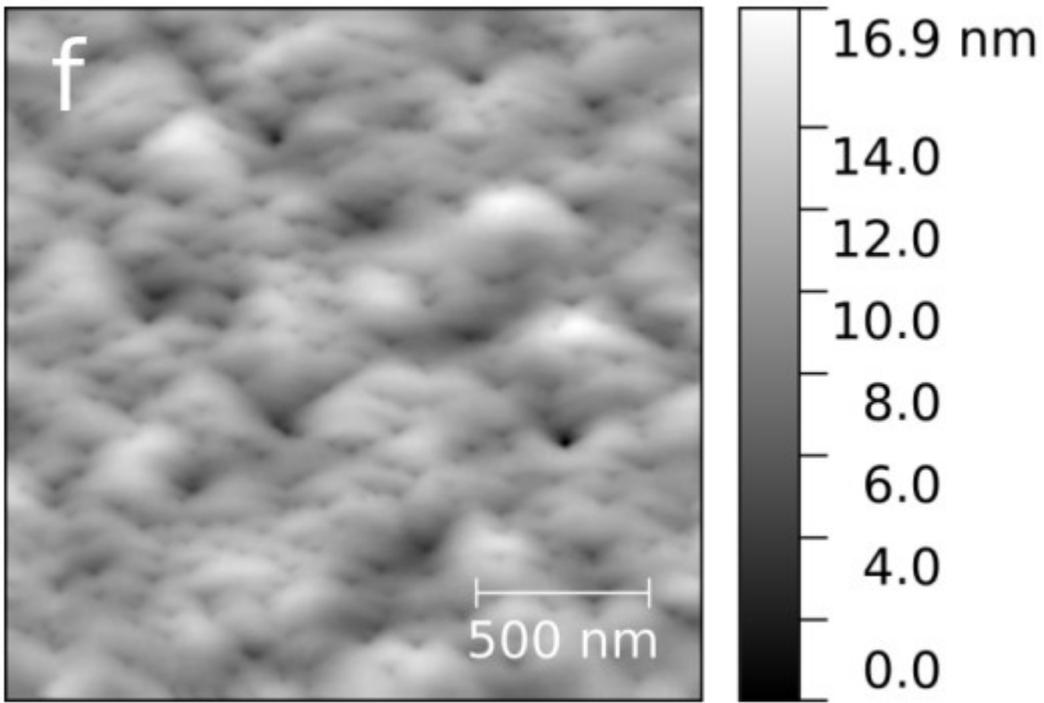

**Figure 2.**

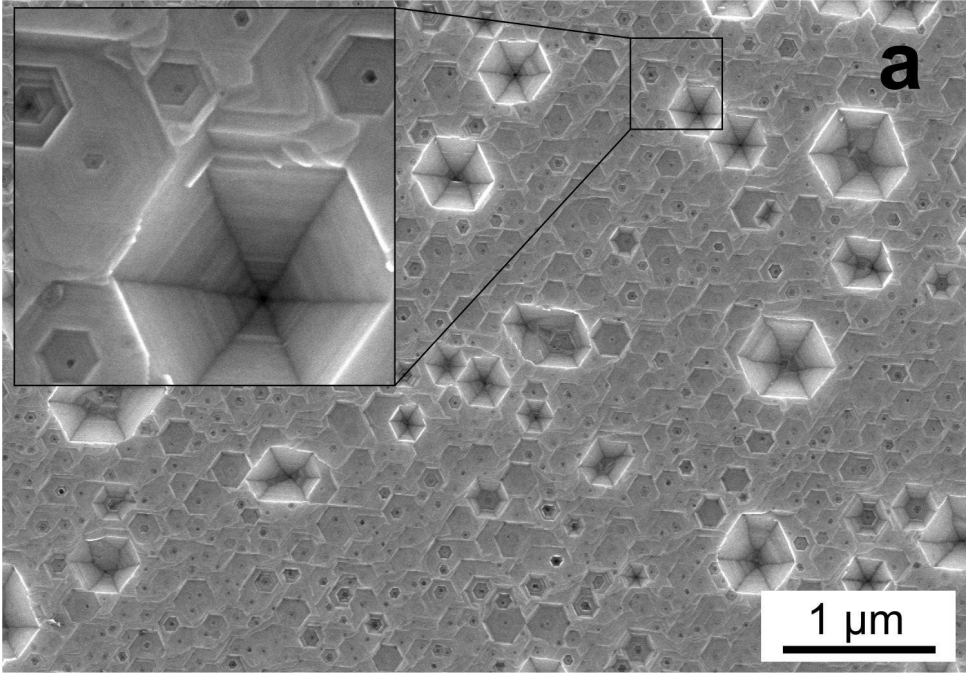
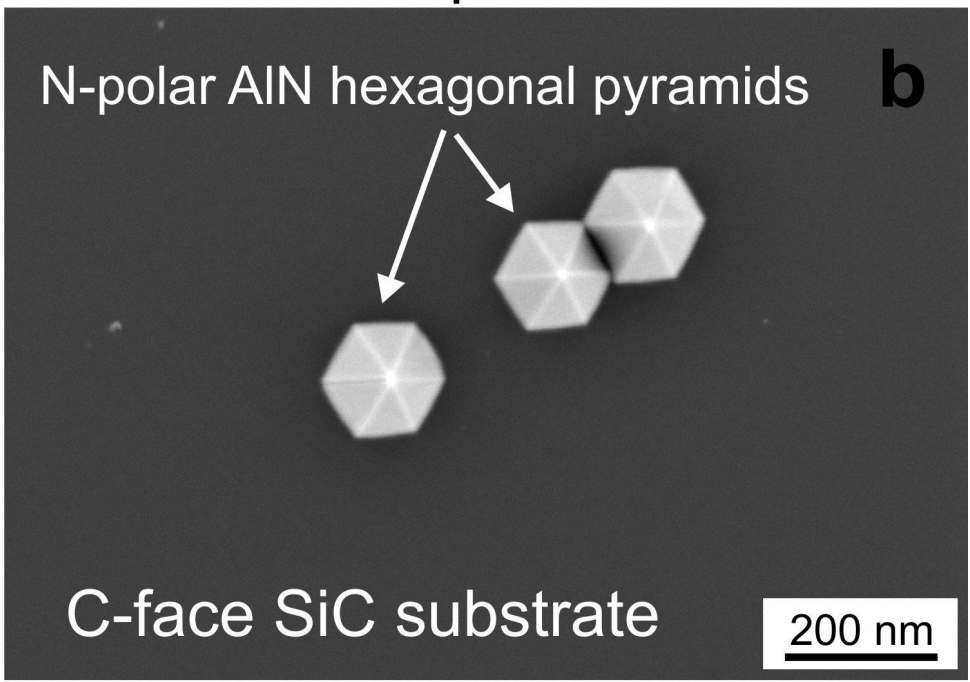
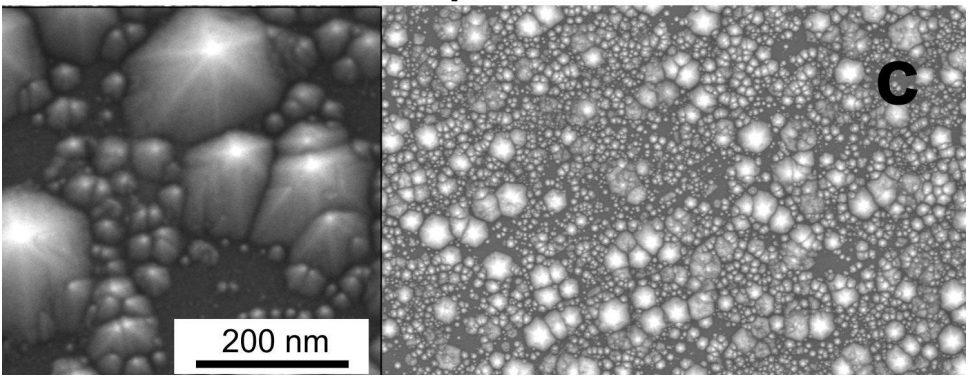

**Figure 3.**

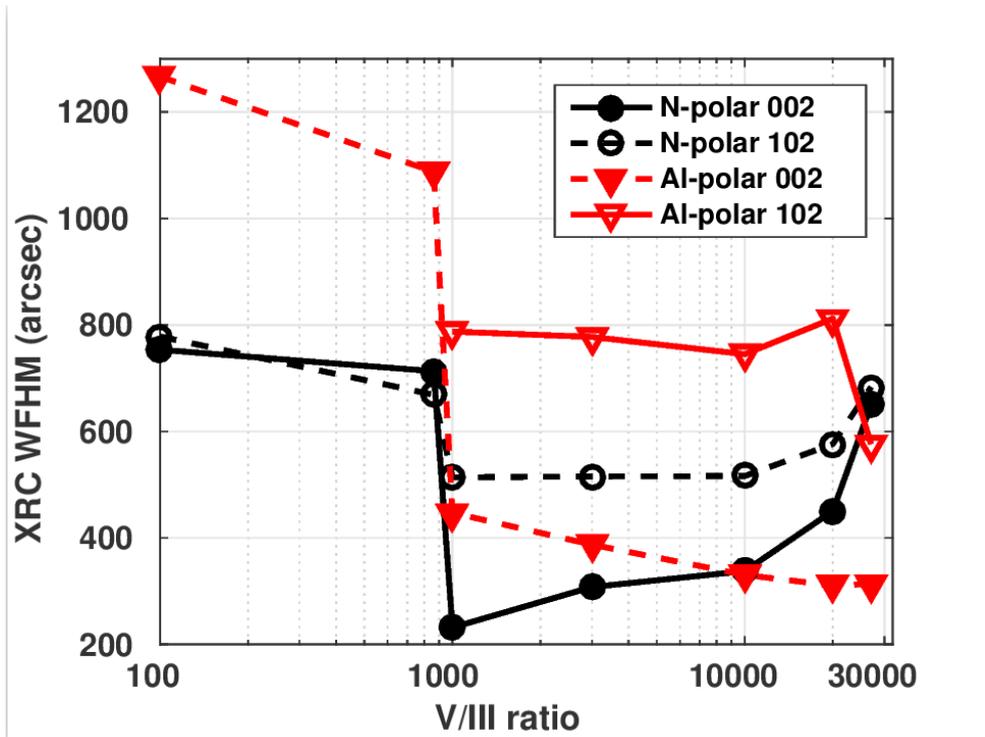

**Figure 4.**

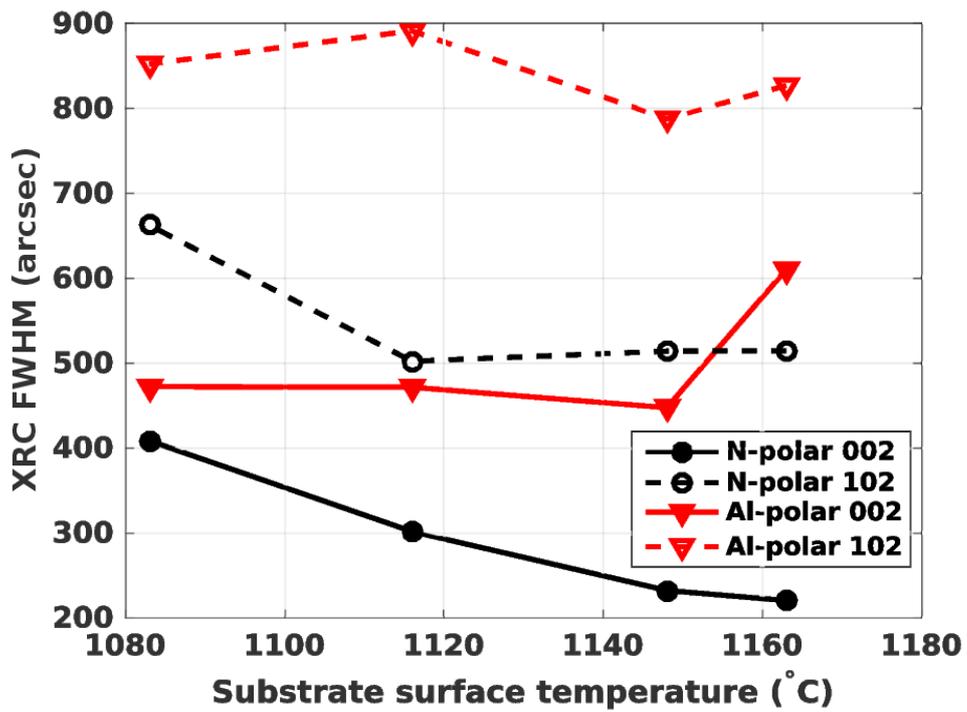

**Figure 5.**

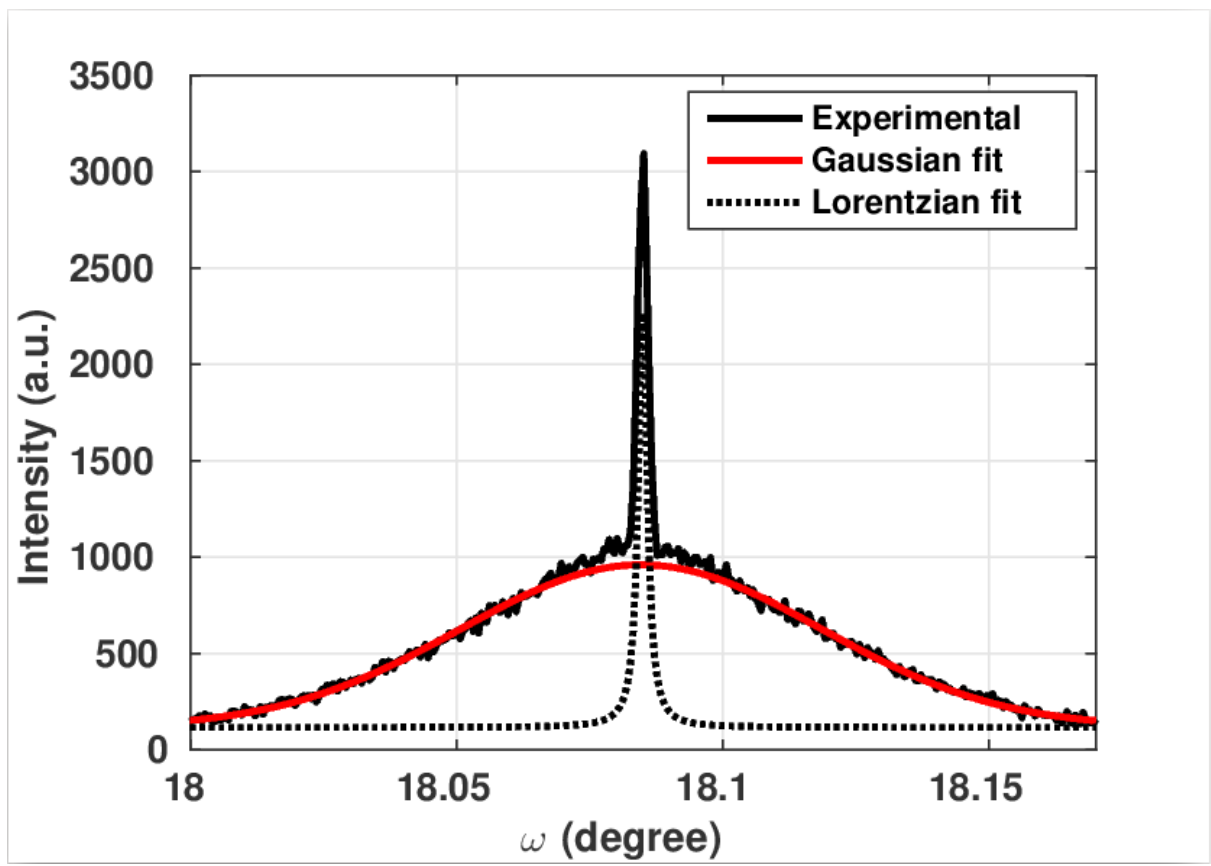

**Figure 6.**

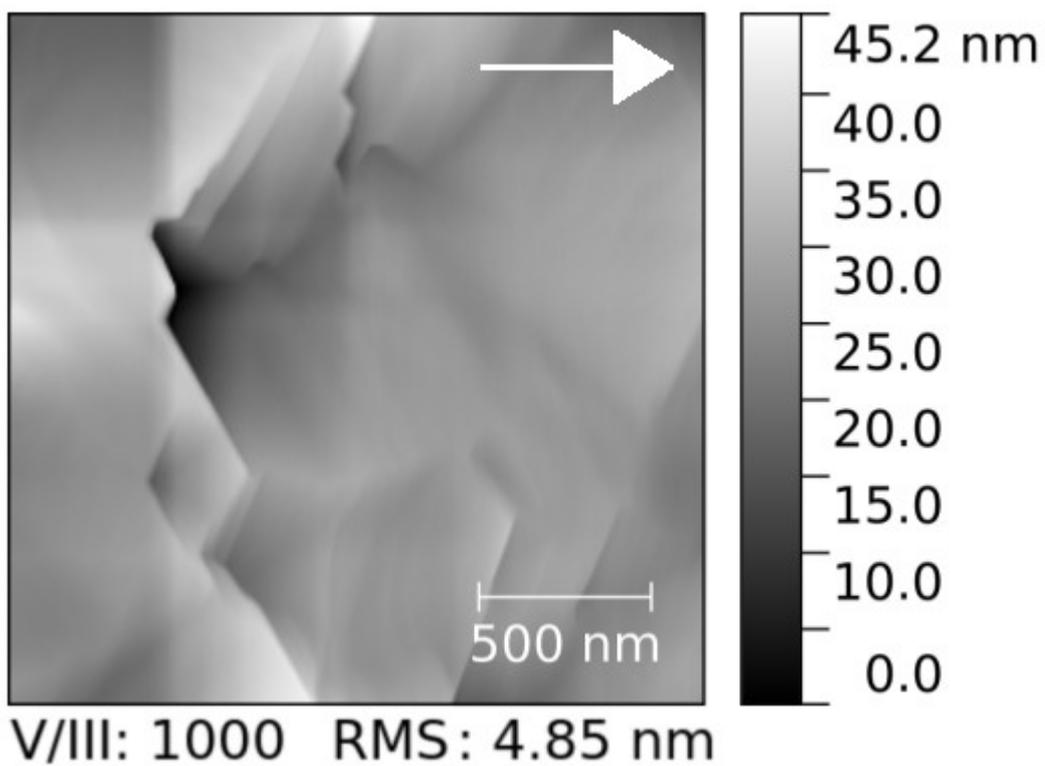

**Figure 7.**

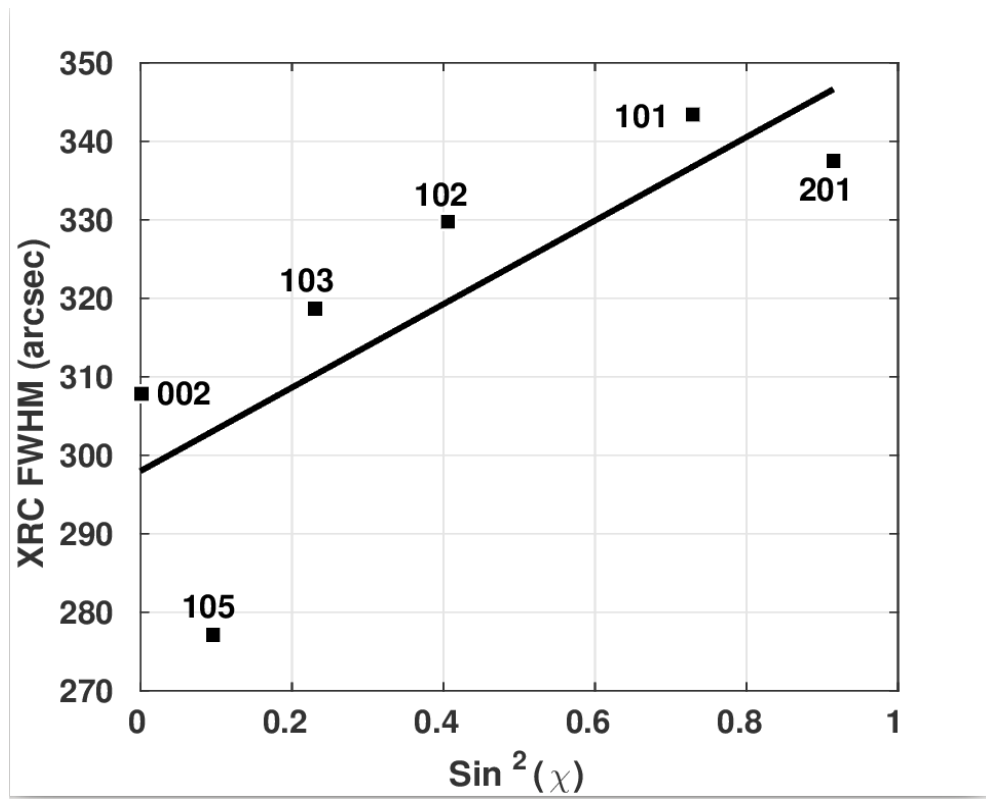

**Figure 8.**